\documentclass[letterpaper, 10 pt, conference]{ieeeconf}  % Comment this line out
\IEEEoverridecommandlockouts                              % This command is only
\usepackage{graphics} % for pdf, bitmapped graphics files
\usepackage{epsfig} % for postscript graphics files
\usepackage{amsmath} % assumes amsmath package installed
\usepackage{amssymb}  % assumes amsmath package installed
\usepackage[ruled,linesnumbered]{algorithm2e}

\title{\LARGE \bf
A Secure Connectivity Model for \\Internet of Things Analytics Service Delivery}

\author{ \parbox{3 in}{\centering Hussain Al-Aqrabi and Richard Hill\\
		 Centre for Industrial Analytics\\
         School of Computing and Engineering\\
         University of Huddersfield, UK\\
         {\tt\{h.alaqrabi, r.hill\}@hud.ac.uk}}
}

\begin{document}
\maketitle
\thispagestyle{empty}
\pagestyle{empty}
\begin{abstract}
Wide scale interest and adoption of Internet of Things (IoT) technologies is fuelling innovation in the way individuals and even machines can interact to exchange knowledge. One area of particular interest is that of analytics. Ever-decreasing form factor hardware is enabling computation and data storage to be embedded into many different devices. The combination of network connectivity and emerging distributed models of service orchestration is allowing the creation of new ways of measuring, monitoring and analysing performance. Using an approach inspired by the NIST seven layer model of cloud computing, we propose a model of connectivity that enables analytics services to be consumed across individual system components that are distributed, such as those found in the IoT and Industrial IoT (IIoT) domains.
\end{abstract}
%
%\providecommand{\keywords}[1]
%{
% \small	
% \textbf{\textit{Keywords---}} #1
%}
%\keywords{cloud computing, distributed systems, security, authentication, trust, multiparty, Internet of Things}


%\end{keywords}
%%%%%%%%%%%%%%%%%%%%%%%%%%%%%%%%%%%%%%%%%%%%%%%%%%%%%%%%%%%%%%%%%%%%%%%%%%%%%%%%
\section{Introduction}
The ubiquity of opportunity offered by the Internet of Things (IoT) is providing new ways to embed computation and storage into everyday devices. Reductions in the cost of hardware, coupled with technological advancements, are enabling a constant stream of innovative new products and services that rely upon IoT architectures. Such developments are particularly evident in the `wearable' and healthcare industries, with parallel innovations occurring in the Industrial Internet of Things (IIoT) \cite{bessis2013,hill2017a}.

Along with the ability to sense environments, conditions and processes, comes the requirement to measure, monitor, analyse and evaluate performance. As such, Business Intelligence (BI) is undergoing a resurgence as consumer demand moves beyond traditional dashboards through forecasting, towards prescriptive analytics. Of late, vendors have delivered BI often as a flexible, on-demand service, making use of the underlying elastic cloud platforms that enterprise software applications are typically deployed upon \cite{alaqrabi2012,alaqrabi2013,alaqrabi2015,alaqrabi2018}.

Advancements in technology have increased the ability to connect such a variety of embedded devices to larger pools of resources such as clouds. Integrating embedded devices and cloud servers raises an important discussion regarding the nature of data generated or transmitted by IoT devices. These approaches must be secure and provide the necessary privacy controls for users. At present, the security and privacy concerns created by these devices play a central role in the successful integration of these two technologies \cite{kalra2015}.

The heterogeneous nature of IoT environments makes it much harder to detect the insider and outsider attacks in such universal platforms \cite{Alrawais2017}.

Experiences with clouds, especially those that have public or hybrid architectures, illustrates that cloud services and applications require multi-layered approaches to external and internal threats to security. This is most pertinent in the IIoT environment, where business systems contain valuable Intellectual Property (IP) that can only be protected by retaining tight security over working practices.

This article describes a model for delivering analytics services to and between components such as those encountered in IoT architectures. A number of adversarial attacks are simulated to demonstrate the effectiveness of a cloud-inspired multi-layer security model in the IoT domain.
\section{Cloud Computing Model}
The core principles of cloud computing: on-demand self-service, broad network access, resource pooling, rapid elasticity and service metering (NIST Special Publication 800-145) \cite{hill2013}, make utility computing an attractive proposition for environments that contain distributed resources.
The NIST definition \cite{nist2011} describes five options for the deployment of cloud computing as follows: public clouds, private clouds hosted onsite, private clouds hosted off-site, onsite community clouds and offsite community clouds.

Abstraction is one of the most compelling motivations for system architects; this enables myriad heterogeneous resources to be viewed as a homogeneous whole. Cloud architectures of any of the five types defined by NIST are modelled using a framework that consists of seven layers. The physical infrastructure foundation is layer one, upon which a resource abstraction (virtualisation) second layer resides. Layer three is the resource composition layer, and layer four refers to the Infrastructure as a Service model (IaaS). Platform as a Service (PaaS) exists on layer five, Software as a Service (SaaS) on layer seven and finally the cloud tenants' applications sit within layer seven \cite{demchenko,hill2013}.

We have considered the application of analytics services in the context of an enterprise Business Intelligence application that is likely to be delivered across a heterogeneous network of devices. In a cloud environment this is typified by off-site private or community clouds, whereby multiple disparate business organisations can host their own enterprise software services and data repositories in what is, to all intents and purposes, a set of hardware resources that is dedicated to each tenant \cite{demchenko2012,hill2013}.
The tenants access their dedicated resources via a VPN, thus maintaining clear separation between corporate services and their respective analytics implementations. However, should any of the businesses wish to share services to promote operational efficiencies, this can be enabled via community-based agreements \cite{demchenko2012}.

We have used the cloud model as inspiration to propose a multi-layer service-oriented framework that can deliver secure services (such as BI analytics) across distributed infrastructure. Security and privacy are key concerns for tenants of shared cloud infrastructure such as that found in off-site private and community clouds. The sharing of services such as malware detection and inoculation has significant benefits for organisations who may not otherwise have strategies or the resources to maintain their own defences. We shall now briefly review services dedicated to security and privacy within clouds in the following section.
%
%The proposed architecture comprises multi-layered security and privacy services offered at a cost, which may be unaffordable to retail Cloud clients. Hence, public Clouds have been kept outside this solution. In an outsourcing environment, each proposed layer is a Cloud in itself offering the specific security or privacy control in a service-oriented framework. For example, the anti-malware layer offers an array of Cloud databases comprising records of signatures and traces of malware entering through compromised sessions (i.e., embedded in the session packets). 
%Before getting into the solution, a review of security and privacy as a service on Cloud computing is presented in the next section.

%
\section{Secure Service Delivery}
In keeping with the service orientation models described so far, security as a service can be considered as a multi-layer framework in itself \cite{panian2008}, protecting each layer of the cloud computing model \cite{carvakho2011}. Such a service lends itself to a utility offering in the same way that clouds are rapidly provisioned and expanded on demand, for a cost that is quantifiable and chargeable to the consumer.
%In this model, every security layer mapped with the Cloud layer may be offered as a chargeable service to the clients [16]. Such a framework may be deployed as per the principles of trustworthy computing at each layer [16].

As described above, resource abstraction through virtualisation is a key principle for a cloud-inspired architecture, and therefore any security and privacy services need to be made available to all relevant Virtual Machines (VM) for each client \cite{kumar2011,luo2011}. 

Therefore, security and privacy control services must be orchestrated via appropriate service interfaces between each cloud and its respective tenants that are accessing the services \cite{kumar2011}. This control will be managed by the relevant virtualisation security manager for each VM, and as a consequence the control must be located within the virtualisation layer.

Privacy is managed via policy resources that need to be retained securely in a protected space such as a digital vault for encryption keys or digital certificates and such like \cite{pearson2009,diaz2013}, all contained within a tier above other layers such as authentication and client metadata layers \cite{pervez2012,chadwick2012}.
%This layer may be added over an authentication layer and a tenants’ metadata layer defining all attributes of the authorized tenants required for accessing appropriate resources on the Cloud [20][21]. 

Any instances whereby session packets do not have the requisite authority to invoke a particular service interaction will result in the session being terminated, and is a key security feature that is embedded within the framework.

%A session can be denied by either of these layers if the tenant-specific information in the session packets does not match the database contents serving these layers. This concept is presented in the form of a detailed multi-layer design in the next section.
%%%%%%%%%%%%%%%%%%%%%%%%%%%%%%%%%%%%%%%%%
%
\section{Multi-layered Security Model}
\begin{figure}[tb]
 \includegraphics[width=\linewidth]{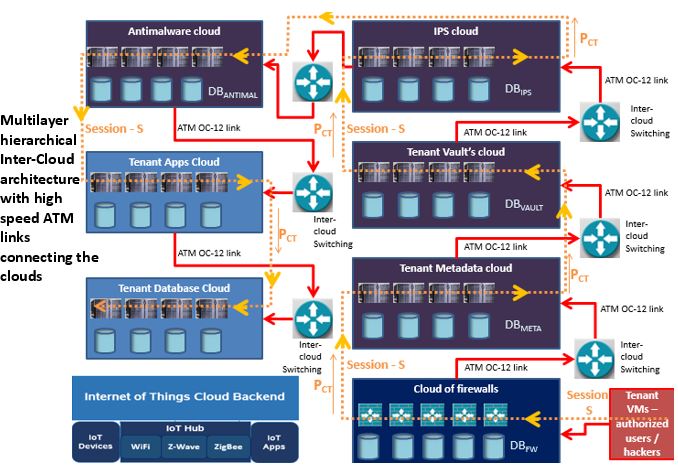}
 \caption{Multilayer hierarchical inter-cloud architecture.}
 \label{fig:multi}
\end{figure}

Our proposed model is illustrated in Figure \ref{fig:multi} and describes the cloud-inspired architecture containing multiple layers. Each layer incorporates security and privacy services that make use of firewalls as gateways for session traffic from each of the respective clients.

We have implemented the model within Opnet, the description of which is as follows.
Each firewall is represented by a Cisco PIX 535, which operates across the various layers including network, transport and application. Access control lists within the firewalls enable the governance of traffic based on IP addresses, protocols and  ports for common and bespoke applications.

The firewalls are configured at the application layer to filter traffic from different URLs, encrypted sessions (HTTPS) and various clients such as Java, and are able to utilise Internetwork Key Exchange (IKE) to encrypt all sessions via DES, 3DES or AES.
% The firewalls are made of Cisco PIX 535 models acting as network, transport, and application layer firewalls. The firewalls can be equipped with access control lists based on protocols, IP addresses, hostnames, TCP/UDP ports for common applications and services, and higher order TCP/UDP ports for custom applications and services.
 %At the application layer, the firewall comprises filters for allowing URLs, HTTPS sessions, FTP sessions, Java clients, and Active X controls, selectively.  In addition, the firewalls can encrypt all sessions using IKE (Internetwork Key Exchange) with IPSec (IP Security) protocol. The encryption services supported are DES (56 bits), 3DES (128 bits), and AES (256 bits).
As an example we have defined four separate LANs to represent different clients/cloud tenants, and they each access the cloud through separate firewalls.

To illustrate an adversarial scenario, we have incorporated a simulated distributed attack from three malicious parties, who each are attempting to access the network through independent firewalls. Such an attack is a key concern for early adopters of IoT technologies, as the pervasive use of wireless communications presents many opportunities for business vulnerabilities to be exposed.

%The tenants are hosted on four tenant LANs accessing the Cloud network through four firewalls.In addition to the tenant LANs, three hackers are shown accessing the Cloud through different firewalls for simulating a distributed attack.
%The hackers have been modelled as clients gaining network and transport level access to the Cloud by breaking the firewalls. They can also be viewed as valid tenants that have gained access to the Cloud by buying a published subscription.
% 
\begin{figure}[h]
 \includegraphics[width=\linewidth]{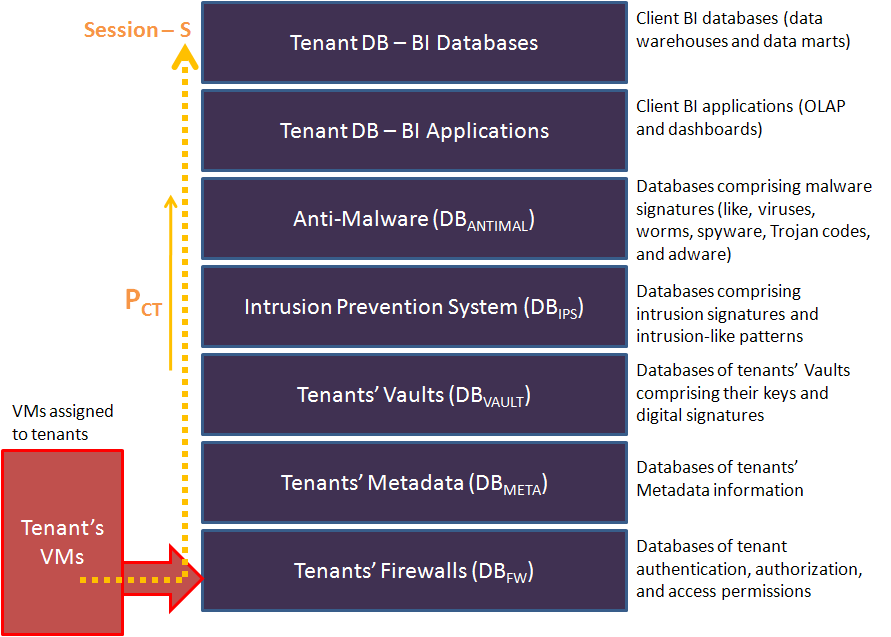}
 \caption{Cloud layers modelled in this work.}
 \label{fig:cloudlay}
\end{figure}

All of the multiple layers of the cloud are illustrated in Figure \ref{fig:cloudlay}. We have embraced the cloud concept of resource abstraction and exploited this in the multi-layer model by representing each of the layers as separate clouds. Within each cloud layer exists an array of computing hardware to maximise performance and therefore minimise response times to service requests. It is an imperative that the model must not introduce excessive overheads into the normal functionality of the system.

The cloud layers are arranged such that inbound traffic from clients are filtered by the firewalls and then passed through successive layers until the system is satisfied that the requests can be delivered to the analytics functionality residing on the cloud applications layer. We now describe each of the layers and the role that they play within the model in turn:

%The tenant sessions enter the Cloud through the firewalls and pass through the layers in a sequential manner. Each layer verifies the sessions and allows further communications with the next layer above it. After a series of examinations, the sessions finally reach the Cloud apps layer running the BI applications. The roles of the layers are explained as the following:
%
%
\begin{itemize}
\item {\bf Tenant firewalls}. This cloud layer consists of a number of databases that hold authentication data for each of the tenants, in order for them to be permitted access to the VM that have been assigned to them as part of their subscription. This is the gateway for each session, $S$ to be invoked.

%Tenants’ firewalls – are deployed for ensuring that authorised tenants are allowed to enter the cloud; includes built-in authentication using RADIUS protocol. This cloud of firewalls comprises an array of databases feeding the firewalls with relevant information. The databases (DBFW) hold the details of the tenant VMs (acquired by tenants through subscriptions on the cloud) and authentication details of the tenants. Once the authentication details are verified (User IDs, passwords, and additional authentication enablers like a physical token card), the session (S) is allowed to proceed to the next layer. 
%
\item {\bf Tenant metadata}. Beyond the authentication criteria that is marshalled by the tenant firewalls layer, there exists further metadata for each client tenant. This metadata describes the credentials to authorise access to specific instances of repositories, applications and services. The detail is embedded within the session packets and is used to verify whether the session can continue or not.
%b)	Tenants’ Metadata – comprises detailed tenant information for ensuring their authorization to destination application instances and database objects. Permissions details are embedded in the tenant session packets (PCT) by the tenant initiating the session (through a form). The metadata (DBMETA) inspects the details and allows the session to proceed if there is a 100% match.
%
\item {\bf Digital vaults}. The vaults are a secure place to retain digital signatures/certificates and decryption keys so that only the requisite authorities can access their own content and services. Public key encryption ensures that legitimate session requests are honoured and bogus requests are terminated.
%c)	Tenants’ Vaults – comprises decryption keys or digital signatures for unlocking the destination application instances and database objects. The tenant need to issue a request public key to obtain the private key needed to encrypt the session for proceeding further. If a wrong key is detected by (DBVAULT), the session is blocked.
%
\item {\bf Intrusion Prevention System}. The occurrence of adversarial attacks is not limited to the correct authentication at the start of a session. Intrusion detection prevents attacks on sessions that are in progress, such as SQL injection in web forms. This might be considered a route into the DB\textsubscript{META} or DB\textsubscript{VAULT} repositories from the perspective of an adversary. This cloud layer prevents such activity from continuing.

%d)	Intrusion Prevention System (IPS) – checks for traces of exploit signatures in the ongoing sessions. A hacker may have executed exploit codes (example, SQL Injection or cross-site scripting) to steal data from DBMETA and DBVAULT. If this is the case, the DBIPS will detect the traces of exploit codes and block the session. 
%
\item {\bf Malware protection}. Similar to the IPS cloud layer, an anti-malware layer contiuously montitors for trojan activity, where malicious exploits are embedded and concealed within session packets, only to be executed at the application layer. Records of the monitoring and detection history are retained within DB\textsubscript{ANTIMAL}.
%e)	Anti-malware – checks for viruses and spyware signatures in the ongoing sessions. The sessions passing through the IPS may have spyware, adware, or worms embedded in the packets for executing at the BI application layer. The database DBANTIMAL will be able to detect and block the session if such malware traces are detected. 
%
\item {\bf Tenant applications}. This layer hosts the tenant's applications themselves, in this case the analytics functionality of enterprise BI. The preceding layers ensure that only a marshalled session, $S$ can access this layer, which in the case of BI potentially provides access to confidential business operations and performance data. For some functionality, there will be a requirement for further authentication from a user, such as an account and password, etc.

%f)	Tenant Apps – comprises suites of BI applications. At this layer, the tenant is able to gain access to the presentation screen of BI comprising dashboards and OLAP query tools. The session S will reach this stage after crossing all the checkpoints in between. There may be an additional level of authentication (BI-level user ID and password) included at this layer.
\end{itemize}
\subsection{Tenant repositories}
The final cloud layer hosts the back-end repositories that serve myriad tenant applications. For an analytics application, this would be the databases/data warehouses/data lakes that retain the underlying business data, together with any processed data objects for reporting and analysis. These objects are typically accessed by tenant users through reporting dashboards and data visualisation suites, abstracting users away from the complexities of database organisation.
\subsection{Simulation design}
%
%The cloud layer services are modeled as shown in Figure 2. Each security service has a built-in relational database for examining the tenant sessions. The databases at the firewall (DBFW) cloud help in authenticating, authorisation and providing access permission to the session S initiated by the tenant. Tenant metadata (DBMETA) comprises tenants’ information for enabling authorisation to designated application instances and database objects that the tenant has subscribed. Tenants’ vault comprises keys and digital signatures stored in database objects (DBVAULT) that can be retrieved based on tenant authorisation enabled by the metadata layer. The IPS comprises a database of known exploit signatures (DBIPS) such that a tenant’s attempt to launch exploits through an authorised session can be matched with them, detected, and blocked. The anti-malware comprises a database of known malware traces (DBANTIMAL) such that a tenant’s attempt to spread malware through authorized sessions can be detected and blocked. The cloud DB comprises the database objects that the tenant is authorized to access. These objects are locked and can be unlocked by the keys picked up from the tenant vault layer.
We have elected to model and simulate the proposed system (Figure \ref{fig:cloudlay}) so that we can examine the operational characteristics in terms of performance, and its ability to provide a collection of services that are resilient towards malicious attacks.

%OPNET does not have the feature to create database tables and enter content in databases. Hence, the simulations are limited to studying their operating behaviour, performance levels, and blocking of attacks.

%The applications are packaged in profiles as shown in Figure 3. In this design, the profiles represent a pool of virtual machines in which, all the security/privacy services, the Cloud BI apps, and the Cloud BI database are packaged. There is no profile configured for allowing access to these applications services outside the virtual machines. All the servers on the Cloud run these virtual machines profiles only.

\begin{figure}[h]
\centering
 \includegraphics[width=0.85\linewidth]{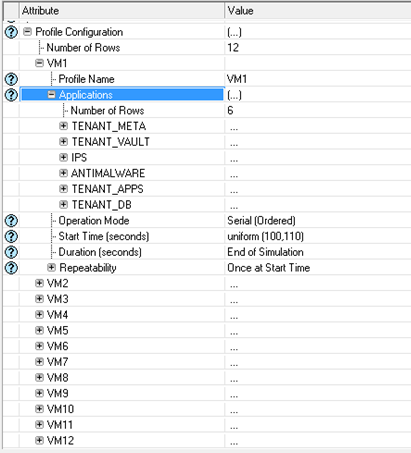}
\caption{Virtual Machines with applications packaged.}
 \label{fig:virmac}
\end{figure}

Figure \ref{fig:virmac} illustrates how the cloud layers have been mapped to individual profiles. Each profile consists of a collection of VMs, that host the contents of the model as described in the previous section, namely: security and privacy services, tenant applications and repositories.
\section{Modelling adversarial attacks}
\begin{figure}[tb]
\centering
 \includegraphics[width=0.9\linewidth]{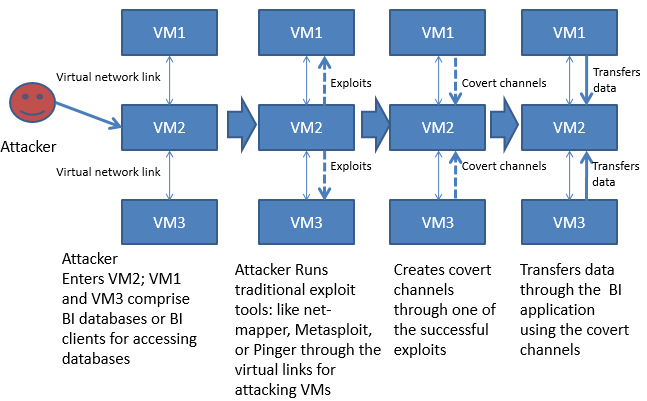}
 \caption{An attack scenario showing the problem.}
 \label{fig:attack}
\end{figure}

If we now consider an adversarial attack upon the model, we can see how the system protects against such a scenario.
Figure \ref{fig:attack} shows the situation where a malicious agent attempts to infiltrate the system to obtain access to an authorised tenant's VM. Whilst the cloud service requires verification to be able to subscribe and access the remote resources, what appears to be a legitimate tenant could actually be an adversary that is masquerading as a valid client, who has the objective of entering the cloud and then attacking other tenant VMs from within the same cloud.

Tools such as Metasploit can be employed to automate the delivery of exploits in a rapid fashion. This could enable a bogus tenant to create surreptitious means of exposing sensitive data, unbeknown to any other party.
 %As per the scenario shown in Figure 4, the attacker may deploy effective exploitation tools like Metasploit and Netmapper on their VMs. Metasploit works in a terminal emulation mode. The attacker may simply have to store its files on the virtual disks on the Cloud accessible through the VM and launch the terminal emulation. It launches an emulation screen and allows the attacker to execute thousands of exploits from its internal database of exploit codes and their payloads. The entire process is quite user-friendly and expert attackers can even modify its codes to make it operative in any environment. The process is described in detail on metasploit.com. It is also used as a penetration-testing tool given its stealth capabilities. Metasploit can help the attacker to penetrate the neighboring VMs and create covert channels through which, data proliferation can be carried out without anyone knowing about it.
% 

\begin{figure}[tb]
 \includegraphics[width=\linewidth]{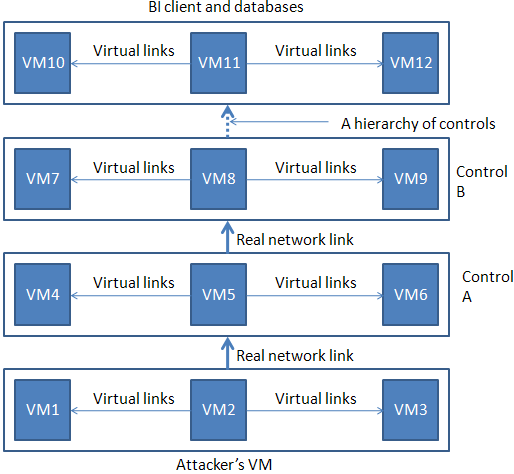}
 \caption{A hierarchical framework presented as a solution to the attack scenario in Figure \ref{fig:attack}.}
 \label{fig:ahier}
\end{figure}

Posing as a legitimate tenant, the adversarial agent would in this case be attacking from a VM that is authorised and hosted within the cloud. As such, the conventional cloud security processes would not be able to detect such activity.
%When such exploits are launched from the VM owned by the attacker, the Cloud security controls will not have any ways to know about the attacks.
In effect, the activity is obscured by the sheer volume of VMs that exist within a cloud environment.
This is a significant challenge for cloud service providers, particularly as service orientation through Microservice Architectures becomes more prevalent \cite{hill2017}. If we extend this to the IoT domain, there is a stronger desire to package functionality into services, to be hosted on distributed, connected hardware. Therefore, the ability to successfully address this issue is a key feature of this work.
%This is because the attacker has already entered the maize of VMs hosting hundreds and thousands of them. The Cloud providers may be having sound peripheral security controls around this maize of VMs for protecting against external attackers. However, what could be done when an attacker has access to a VM deep inside the maize by merely buying a premium subscription? There needs to be a solution to such a scenario.
We propose a solution whereby the collection of VMs are organised into a hierarchy, as Figure \ref{fig:ahier} shows.
%The solution can be implemented by converting the maize of VMs into a hierarchical framework as presented in Figure 5.
If an adversarial agent enters the cloud via a subscription, and is assigned VM2, it has the potential to employ cross-channel attacks against VM1 and VM3, thereby exploiting the presence of virtual links between differing VMs. Typically, cloud security controls are deployed to prevent external attacks rather than insider attacking.
%In the modified scenario, the attacker is shown as entered VM2 through a formal subscribing process of the Cloud service provider (perhaps, paying the highest subscription and getting a premium user status). The attacker can attack VM1 and VM3 through cross-channel attacks (VM to VM) because the virtual links are insecure. The security controls are normally deployed outside the hosts having VMs and hence they can protect against external attackers only. In this scenario, the attacker has gained access to VM2 and hence is an insider attacker. This might have happened through subscription validations.
% 
This architecture limits the opportunity to commit further exploits since the attacker is prevented from moving to the next control using virtual links. The only option is to proceed using a real network link by requesting a session in Control A, to communicate with VMs 4,5 and 6. Since there now exists Control A, the attacker has to successfully satisfy {\tt Tenant Metadata Inspection} in order to proceed with the penetration.
%However, in this architecture the attacker will not gain anything by attacking VMs 1 and 3 because they have nothing except a BI client that can proceed only through the next control (example, Tenant Metadata inspection). Hence, the attacker will have no choice but to proceed through the Real network link and establish session with VMs 4 to 6. These VMs hold the first control (tenant metadata inspection).
Of course, an orchestrated and sustained attempt to commit an attack will mean that we should anticipate an adversarial agent will also have obtained valid credentials, either by posing as a legitimate tenant or otherwise.
%The attacker may cross this control if he/she has genuine tenant credentials. 
In this case, Control B will have required that the malicious agent would need to navigate the entire stack of cloud layers before access could be gained to the analytics interface.
%However, the attacker will be countered by the second control in VMs 7, 8, 9. In this way, the attacker will have to breach all control layers successfully before reach the VMs hosting the BI databases.

\begin{figure}[tb]%[h]
 \includegraphics[width=\linewidth]{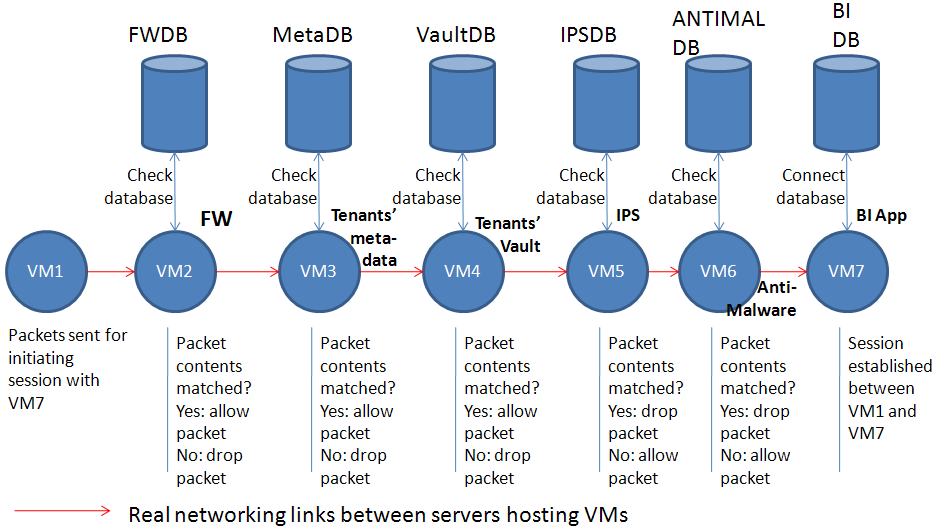}
 \caption{The position of the controls at each layer of the hierarchy.}
 \label{fig:control}
\end{figure}

If the attacker has satisfied the cloud validation and metadata inspection layers, the only way forward now is to plant exploits in the hope that these will lie undetected. However, the Intrusion Protection layer, and the Anti-Malware layer both offer protection for subversive, covert attacks from the inside.
%In order to breach the databases for stealing data of other tenants, the attacker will have to use exploit tools (like, Metasploit, NetMapper or Pinger). However, these exploits will be detected by the IPS and Anti-malware layers. Hence, the attacker may fail to steal any data in spite of gaining access to Cloud VMs by buying subscriptions. For example, this design can protect Amazon EC2 Cloud that has been tested to be vulnerable to VM to VM cross channel attacks.
The result is that our proposed model prevents data breaches, even when adversarial attacks are launched from what appears to be genuine service subscribers.
We can see in Figure \ref{fig:control} a sequence in which various security controls might be instantiated. The security policies of the host system (or systems) will inform the order in which controls are implemented, to suit the goals desired by the infrastructure provider. It is also evident how a tenant's session is routed through the various VMs in order to access the relevant analytics services.
%presents the mechanism in which the controls may be deployed. The sequencing of controls may merely be security policy decision for making the system as effective as possible. The concept will remain the same irrespective of how the controls are organized. Figure \ref{fig:control} also shows the path of the session of a tenant travelling through multiple VMs before reaching the BI application and its databases.
% 
\section{Session packet inspection}
In this section we shall describe a detailed walkthrough of the security model and explain the use of session packet inspection.

Referring back to Figure \ref{fig:ahier}, VM1 will host a client that ultimately will access the enterprise analytics application hosted in VM7. A malicious agent that has access to either VM2 or VM3 (or both) can only attack the client of the analytics application, rather than the application itself.
%The VM1 comprises a lightweight client of the BI application loaded in VM7. The client knows nothing except that it needs to connect to VM7 to proceed further. Hence, an attacker in an adjacent VM will gain nothing by attacking this VM. The attacker will have this client anyways on his/her native VM.
VM2 will be used to test the validity of the VM1 client using a VM identification number, and assuming that all is well, will launch a form via VM3 to request details from the tenant.
%The client may be viewed as an empty window in which, the subsequent controls will launch screens for capturing details. At VM2, the client session will be tested for the source VM (by virtue of a VM identification code) and a screen will be launched for entering authentication details.
The details requested from the tenant will vary each time that VM3 is executed, but they will always refer to some personal details that can be used to help identify the correct tenant.
%At VM3, the client will launch a screen for capturing the tenant details. It may be in a form for randomly asked questions and answers. For example, the form may ask to enter mobile number, social security code, and zip code to proceed further. The questions may vary in each new session.
Once the form on VM3 has been completed, VM2 will use the responses to verify the details against those held in the {\tt MetaDB} repository. Once the metadata has been verified, the session can continue. If the malicious tenant has the intention of acquiring metadata about other tenants, it would need to deliver an exploit into {\tt MetaDB}. Whilst it might be expected that the system is fully patched, it is still conceivable that a vulnerability exists, enabling an adversary to progress to the next cloud layer.
%The tenant cannot move further without answering the questions. The VM2 will check the answers in MetaDB and allow the session to proceed. At this stage, the tenant cannot forge answers unless an exploit is used to penetrate the MetaDB and steal information about other tenants. The MetaDB is expected to be secured and patched with latest updates. However, if the attacker is successful the session can proceed to the next level.
Since a session cannot be interrupted, any evidence left by malicious exploits will still remain as the agent will not be able to remove the incriminating evidence of the exploit. As such, the attacker can only proceed to the next layer by exposing that an exploit has been used to obtain entry. If the malicious agent has obtained private keys, they still cannot progress without exposing details of the exploit within the session.
%It may be noted that the exploit traces have entered the session packets and hence attacker will have no means to clean them before proceeding. This is because the attacker cannot interrupt the session and inspect the packets of the ongoing session.
%A session interrupter needs to be placed between the VMs 4 and 5, which is not there by design. These VMs security controls accessible only to the security administration team of the cloud service provider. Hence, assuming that the attacker is able to breach VM4 to steal digital signatures or private keys, as well, the traces of exploit will be detected at VMs 5 and 6.
Layers 5 and 6 both have the capability to detect malware, which of course is reliant upon adequate, proactive security maintenance to ensure that all exploit databases are current.
%Both the layers are equipped with databases of all known exploit and malware signatures. Hence, the sessions cannot escape them.

It is feasible that an attacker could compromise VMs 3 and 4 with a fresh exploit that has yet to be discovered and documented, in which case the anti-malware cloud layer will be unaware of this exploit as well. Whilst this situation may foil malware detection, there still exists Intrusion Prevention within the cloud layer, which by its nature monitors and reports upon anomalies.
%The only scenario of their success can be the zero day attack when a new form of exploit or malware has been used to penetrate the VMs 3 and 4 and the databases of VMs 5 and 6 are not updated with their information. Now-a-days, security vendors are working actively to counter zero day attacks through their IPS or other form of solutions. A study of websites of security companies like Symantec, Cisco, Trend Micro, etc. will reveal about their research and new innovations. These companies are actively designing their solutions for virtualization platforms.

Our model enables system architects the ability to add or subtract security controls as required for a given set of policies, merely by specifying additional cloud layers for the hierarchy. When a session is authenticated as satisfying the requirements of each layer, it can progress to subsequent layers until the destination application layer is reached.
%This architecture is not about the capabilities of IPS and antimalware controls albeit is about how these controls can be deployed in a multi-layer hierarchy. In fact, any form of new controls can be added to this solution by simply adding a cloud layer. Once the session screen passes all the controls, it reaches the BI App VM where the screen for launching reports and dashboards will be displayed.
One advantage of the use of multiple VMs is that much of the computation can be performed in parallel, and therefore the majority of packet inspection incurs a minimal overhead in VMs 2,3 and 4. However, session packet inspection across cloud layers, especially anything that involves Intrusion Prevention or anti-malware detection will result in an additional overhead. 
%Given that all these layers will have multiple VMs with parallel computing, the inspections can be completed within a few seconds thus ensuring acceptable performance. However, it needs to be considered that the multi-level inspections of session packets will induce delays. While the VMs 2, 3, and 4 will not inspect the packets of an allowed session, the IPS and antimalware in VMs 5 and 6 will continue to inspect each packet passing through them.
%
\subsection{Mapping controls to the seven layer model}
\begin{figure*}[htb]
\centering
\includegraphics[width=0.9\textwidth]{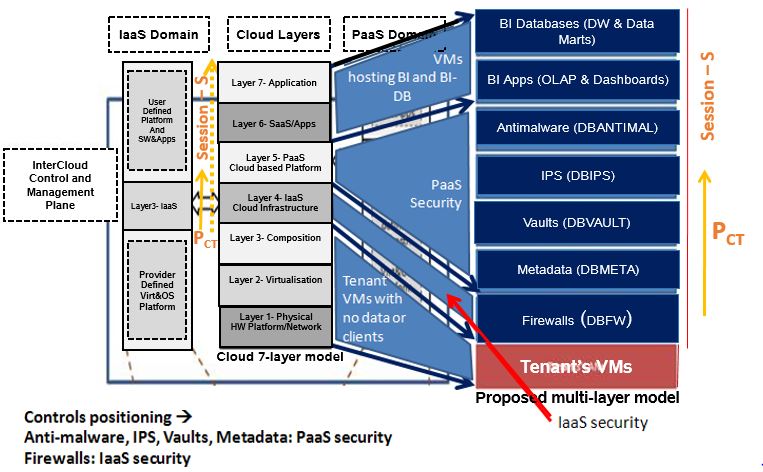}
 \caption{Mapping of the proposed architecture with the NIST seven layer cloud model\cite{nist2011}.}
 \label{fig:mapp}
\end{figure*}

With reference back to the NIST seven layer model \cite{nist2011}, the controls of our proposed model can be mapped to the PaaS layer as per Figure \ref{fig:mapp}, with the exception of the firewalls which naturally reside in the IaaS layer \cite{hill2013,carvakho2011}.
%All these controls will be software-based controls with active databases supporting them. They are expected to act at the platform-as-a-service layer except the firewall, which is expected to act at the infrastructure-as-a-service layer. The mapping of these controls with the seven-layer cloud model presented by references [14] and [15] is presented in Figure 7.
Tenant VMs are hosted in layers 1-3 of the NIST model; users interact directly with clients that do not store data.
%The Figure 7 shows that the layers 1, 2, and 3 of the cloud seven-layer model will comprise tenant VMs with BI clients and no data stored.
%\caption{Mapping of the proposed architecture with the seven layer}
 %\label{fig:mapp}
Sessions commence within tenant VMs, before moving to the application in layer 7 via several layers of verification and authentication in layers 4 and 5. If the tenant has a SaaS platform, this will be made available through layer 6, otherwise all bespoke applications are resident in layer 7.
%In the cloud 7-layer model, the sessions will begin from tenant VMs at Layer 3 (composition) and move upwards up to layer 7 (tenants’ applications). The sessions from the tenant VMs will go through multiple validations in layers 4 and 5 before reaching layer 6 for accessing the SaaS BI application or layer 7 for accessing tenants’ customized applications and databases.
%
\subsection{Session flow}
%
%The flow of session $S$ is shown in Figures 1 and 2, and the validations are shown in Figure 6. The validations are sequential with the positioning of controls shown in Figures 1, 2, and 3, and explained in relation with these figures.
%********************
The proposed architecture is located on layers 4 (IaaS) and 5 (PaaS) of the NIST seven-layer model. All firewalls are categorised as IaaS due to the functionality of verifying VM instance IDs, and also relating these checks to the authorisation data provided by a tenant to secure access to the cloud.
%The firewall is treated as an IaaS control because it will authorize sessions by inspecting VM IDs (in addition to validating the authentication and authorisation details of the tenant for entering the cloud) and hence is related with the virtualization and composition layers.
Whilst VM IDs are assigned in layers 2 and 3, access controls are assigned at layer 4. Subsequent controls are assigned outside of the VM layer, and are primarily concerned with session packet inspection, for a given session $S$.
%The VM IDs will be assigned automatically at layers 2 and 3 and the access controls for them will be assigned at the layer 4 in the firewall.
%The remaining controls are not concerned with the VM layer albeit are deployed for inspecting packets flowing through the session S initiated by the tenant.
Information is requested directly from the tenant to satisfy the DB\textsubscript{META} and DB\textsubscript{VAULT} checkpoint controls, and is supplemented by DB\textsubscript{IPS} and DB\textsubscript{ANTIMAL} controls that perform the session packet inspection function. DB\textsubscript{IPS} and DB\textsubscript{META} are therefore categorised as PaaS controls.
%At DBMETA and DBVAULT checkpoints, additional information is requested from the tenant through forms. The DBIPS and DBANTIMAL are controls for inspecting traces of attack and malware signatures, respectively.
%Hence, they have been positioned as platform controls.
It is likely that controls will also exist at the application layer. For instance, a SaaS instance will require user authentication, as will a custom enterprise application\cite{ouf2011}. User role profiles are useful in such scenarios to manage different levels of system access within an enterprise application, to reflect the role, responsibilities and authority of a particular stakeholder.

%There may be application-level controls as well, which are not shown in this architecture. For example, the BI apps may have own authentication layer. In addition, the tenants may have an additional layer of authentication layer established for accessing data based on their organizational roles. For example, the CEOs may have different access levels than the employees. 

% 
\section{Algorithm design}
Algorithm 1 describes the sequence of inspections that are performed within the proposed multi-layer security model. This algorithm represents the core functionality and can be augmented as additional cloud layers are defined in response to the security needs of an organisation. In the case of a single enterprise the model may tend towards fewer augmentations.
However, enterprises that collaborate, or who choose to adopt shared services across distributed platforms, will no doubt adopt additional layers in order to securely manage service access. 
%The proposed solution is presented in the form of the following algorithm. It is designed to represent the sequence of inspections carried out in the proposed hierarchical multi-layer inter-cloud security architecture. The algorithm can be modified if the security policy of sequencing the controls is changed. However, the basic structure of the algorithm will remain the same. \\
%
% Sample algorithm
%
\begin{algorithm}
\SetAlgoLined
%\KwResult{Write here the result}
Input: $S$, $P_{CT}$, ($DB_{FW}$, $DB_{META}$, $DB_{VAULT}$, $DB_{IPS}$, $DB_{ANTIMAL}$), 1=permit, 0=deny\\
Tenant session: $S$\\
Contents of session packets:$P_{CT}$\\
Contents of FW: $DB_{FW}$\\
Contents of $TENANT_{META}: DB_{META}$\\
Contents of $TENANT_{VAULT}: DB_{VAULT}$\\
Contents of $IPS: DB_{IPS}$\\
Contents of $ANTIMALWARE: DB_{ANTIMAL}$\\
Flags: 1=permit, 0=deny\\
 
 Initialise $S$\;
 Set $S=1$, Match$(P_{CT})$;\\
 \ForEach{\{$DB_{FW}$, $DB_{META}$, $DB_{VAULT}$, $DB_{IPS}$, $DB_{ANTIMAL}$\}} {%
  \eIf{$P_{CT}\in$ \{$DB_{FW}$,$DB_{META}$, $DB_{VAULT}$\} AND $P_{CT}\notin$ \{$DB_{IPS}$,$DB_{ANTIMAL}$\}}{
  set $S=1$; AuthoriseTenantAccess()\; //tenant access authorised\;}{
  set $S=0$; DenyTenantAccess()\; //tenant access denied\;
  }
}
 Output: $S$
 \caption{Multi-layer hierarchical packet inspection}
\end{algorithm}
\subsection{Security model logic}
Our proposed model seeks to address security concerns in distributed service applications across heterogeneous hardware resources by enabling session packet inspection to take place at a number of checkpoints. Each session packet is scrutinised and compared with a number of repositories such as DBMETA, DBVAULT, etc. Each inspection stage shall now be described in turn.
%The contents of the session packets (PCT) are matched with the contents of the databases positioned at each checkpoint (DBFW, DBMETA, DBVAULT, DBIPS, and DBANTIMAL).
%
\begin{enumerate}
\item {\bf DB\textsubscript{FW}:} For each instance, a client initiates a session, which is inspected as a second stage. This session will possess a VM ID together with authentication and authorisation verification data.
%The session is initiated by the client machine (through a browser based VM) at the first step. The first level of inspection is at the second step. The session should contain VM ID provided by the cloud service provider, in addition to the authentication and authorisation checks at the second step.
In addition, the DB\textsubscript{FW} must have an entry that relates to the VM ID, otherwise the session is terminated.
%There should be an entry of the VM ID in the firewall database (DBFW), else the session will be dropped.
\item {\bf DB\textsubscript{META}:} A further inspection is then performed to confirm tenant metadata as requested by the cloud host administrators.
%The next checkpoint is at the third step. The session should comprise tenants’ additional information (like, unique client ID, company name, contact numbers, addresses, landline phone numbers, mobile numbers, e-mail ID, security secrets, and such other identifiers assigned by the cloud administrators). The tenants will be requested for these information through a secure form launched by the metadata inspector.
Once the metadata has been provided, this will be added to session packets. The session can only continue if the session metadata matches that which exists in the DB\textsubscript{META} repository.
%On entering the information, the metadata information will be appended in the session packets that should match the registries in the tenant metadata (DBMETA). The session is allowed only if the information contained in the session packets match the registry entries.
\item {\bf DB\textsubscript{VAULT}:} At this stage the VM ID within the session will be inspected to verify that a private key exists within the DB\textsubscript{VAULT} so that encrypted databases can be accessed.
\item {\bf DB\textsubscript{IPS}:} On reaching this stage, the session itself is now regarded as being fully authenticated and is authorised to progress to subsequent layers. The next stage is to inspect the sessions for evidence of potential exploits that match those held in the DB\textsubscript{IPS} repository. If a match occurs, the session is terminated.
\item {\bf DB\textsubscript{ANTIMAL}:} This stage performs an anti-malware check, which in conjunction with the DB\textsubscript{IPS} inspection, prevents fully authenticated VM sessions being able to penetrate the upper layers of the security model by masquerading as legitimate cloud tenants. 
\end{enumerate}
The algorithm ensures that after a session is initiated, it is inspected at each layer, where each layer is represented by a separate cloud. A key principle is that session packet data must match the firewall data, tenant metadata and vault data before a session can be authorised. Furthermore, the session cannot be granted access to application layers until it has been successfully screened against IPS and anti-malware repositories.
\subsection{Implementing the model}
For our example scenario, the LAN contains 500 clients. Each of the clients is assigned three VMs, with an assigned destination being the tenant client's metadata repositories rather than the eventual analytics application. This was a conscious decision to prohibit any tenant sessions from attempting to subvert the security and privacy controls of the cloud.
%This LAN has 500 client machines and three VMs are allocated for each client named – VM1, VM2, and VM3. It may be noted that the destination settings on this LAN is the Tenants’ Metadata servers, and not the main BI applications and databases. This has been configured to ensure that the tenant sessions do not bypass the first security/privacy layer of the cloud.
For a physical network, this control would most likely be enacted by a firewall in an application layer, or as part of a setting in a virtual network controller. For consistency we have replicated this by ensuring that the destinations of the metadata servers are directed towards the tenant vaults; this is a faithful representation of the intentions of the algorithm, in that it is governing the marshalling of each session packet by enforcing checkpoints in a particular sequence. In the case of a session packet not containing a tenant key that matches a corresponding entry in DB\textsubscript{VAULT}, the encryption key is not assigned and the packets are dropped \cite{diaz2013}. We have represented the adversarial agents as clients who have the authorisation of a valid tenant, in order to simulate an attack from within.
\section{Results}
\begin{figure}[tb]%[h]
\centering
  \includegraphics[width=0.9
  \linewidth]{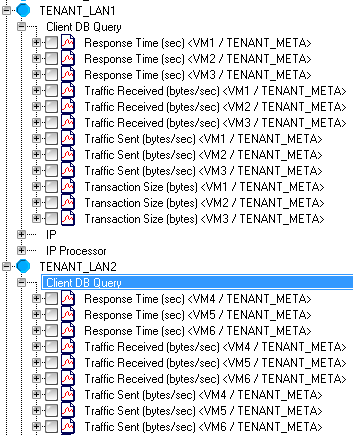}
\caption{Client DB sessions on Tenants' LAN.}
  \label{fig:dbsess}
\end{figure}

%
%The results are summarised in ?????
Figure \ref{fig:dbsess} shows the sessions that have been hosted by the tenant LAN in the simulation. We observed that sessions only existed within the tenant LAN for tenants that had matching metadata in DB\textsubscript{META}. It follows that the sessions that were invoked within the tenant LAN, had consistency between tenant metadata (DB\textsubscript{META}), tenant vault (DB\textsubscript{VAULT}), and the associated VM ID from the initial authentication.
%presents the sessions encountered by the tenant LAN. It may be observed that all the sessions initiated from the tenant LAN are with tenant Metadata only through all the virtual machines.
%Similarly, the sessions are between tenant Metadata and tenant Vault only for the same virtual machines.
The simulation demonstrated that these validations were maintained throughout the experiment for all network hops, illustrating that VMs were complying with their specified destination configurations, and there was no evidence of VMs by-passing authentication layers. As such, all security and privacy control rules are enforced by the model.
%This has been observed for all the hops designed in the model. The results indicate that the virtual machines cannot jump a layer given their pre-defined destination preferences. In this way, the session inspections and forwarding/dropping are mandatorily implied on each VM.

%
\begin{figure}[tb]%[h]
 \includegraphics[width=\linewidth]{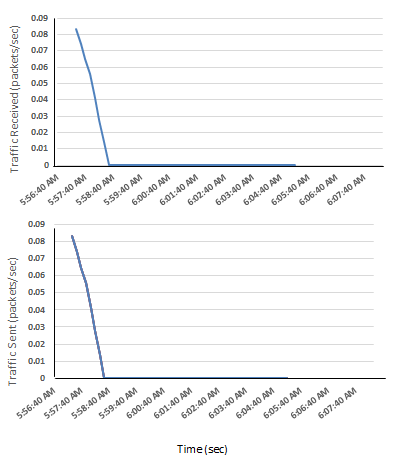}
 \caption{After initial attempts to penetrate the system, IP packets from the attackers’ machines are dropped.}
 \label{fig:ipdropped}
\end{figure}

\begin{figure}[tb]%[h]
 \includegraphics[width=\linewidth]{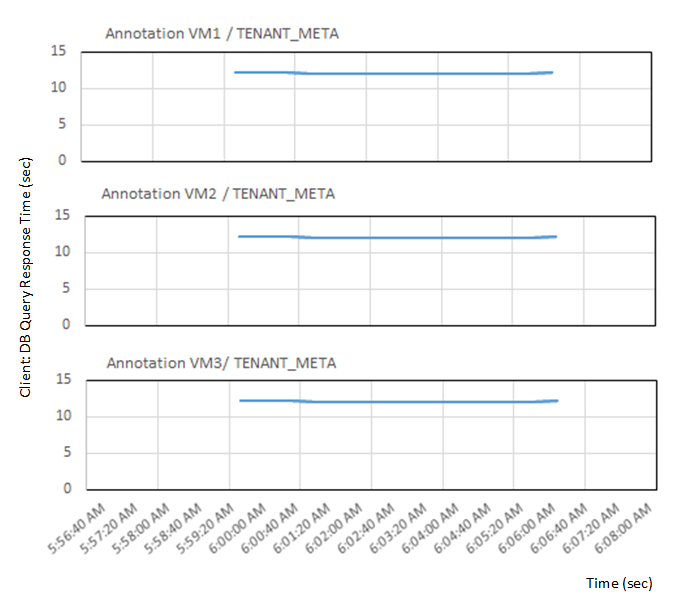}
 \caption{Authorised tenant LANs established, DB sessions initiated with the TENANT METADATA.}
 \label{fig:auth}
\end{figure}

% done
VMs initiated by malicious agents have separate profiles from those of legitimate tenants, even though a malicious agent may appear legitimate at the outset. This means that adversaries had distinct metadata and decryption keys in their vault repositories.
We can see in Figure \ref{fig:ipdropped} the instances where a malicious agent's packets have been dropped as a result of packet inspection through either the IPS or anti-malware layers, preventing deeper penetration into the system. As discussed earlier, an internal attack (an agent with legitimate DB\textsubscript{META} data) would need to compromise DB\textsubscript{VAULT}, DB\textsubscript{IPS} and DB\textsubscript{ANTIMAL} in sequence if it is to successfully reach the analytics application layer.
%All VMs assigned to the adversaries remain separate from the tenant metadata and vault application profiles.
%Figure X shows the packets that are dropped after an adversarial agent has attempted to progress further into the system. In order to succeed with an attack, it would be necessary to breach the vaults, IPS and anti-malware layers, assuming that the hacker had sufficient credentials to satisfy the metadata layer.
% done
%The VMs assigned to the hackers are kept out of the tenant Metadata and the tenant Vault application profiles. In practice, this scenario may be viewed as the unauthorized users not having any records in the metadata or the vault when trying to access a different domain than allowed to them. It may be observed in Figure 8 that the IP packets from the hackers’ LAN dropped after an initial attempt.
% done
Conversely, an authorised tenant may elect to execute DB sessions on their own LAN, since the application profiles can include references to their VMs, as per Figure \ref{fig:auth}. In such cases, the sessions are authorised in the sense that they fulfil the relevant rule in Algorithm 1. Since we have chosen to model each of the layers as clouds, each of the multiple layers can in fact be serviced by different cloud providers. This architecture thus demonstrates significant flexibility and is attractive to system architects who are proactively designing systems that will rely upon heterogeneous hardware and distributed resources, such as the IoT and IIoT environments.
%On the other hand, the authorized tenants’ LANs could run DB sessions throughout the simulation period as shown in Figure 10.
%This is because their VMs are added to the application profiles of all the layers (i.e., their sessions fulfill the rule shown in the algorithm).
%This is a multi-layer security architecture in which, each layer can be a Cloud in itself and served by a different Cloud service provider. Hence, this entire architecture is termed as a multi-layer security as a service framework for tenant organizations hosting BI applications and databases on outsourced private and community Clouds.
An enterprise may adopt this security model so that it can take the opportunity to employ software applications and services that are themselves hosted on offsite private or community clouds. However, the use of these services may increase operational costs, although this is off-set by a reduction in capital expenditure. Operational costs may also increase indirectly through the maintenance charges associated with managing the database updates of five security control layers.
%These services may be premium and expensive and hence, may not be suitable for public Clouds. In addition, there are challenges of maintaining five inspection-aiding databases for keeping them highly performing and up-to-date. There may be costs charged to tenants for records maintenance in tenants’ firewalls, tenants’ metadata registries, and tenants’ vaults. All access permissions may be based on the Unique VM IDs of the browser-based VM access allocated to the client. The IPS and antimalware databases can be updated regularly through their websites of original software manufacturers (OSMs). Given the volumes expected in these Clouds, such updating will require inter-Cloud communications such that all databases in the IPS and Antimalware arrays can pull records from OSMs databases directly. 
% 
%
%%The databases are intended to be hosted in a parallel fashio as per the model. In practice the total number of servers will be much larger than the example.
%All the five databases should be hosted on server arrays with parallel processing, as configured in the model. The number of servers per array will be in hundreds and not the few shown in the model. Hence, this model is designed to establish the concept only and not demonstrate the expected volumes.
%
\section {Conclusions}% done
This article proposes a multi-layer hierarchical inter-cloud security model, that is inspired by the NIST seven-layer model of cloud computing. By using sequential session packet inspection techniques, we have demonstrated an architecture that exhibits considerable resilience towards both external attack as well as more surreptitious internal adversarial behaviour. Whilst VM vulnerabilities are well documented in multi-tenant shared environments, our five layers for packet inspection enables the architecture to identify and compartmentalise malicious activity. Thus, penetration of a firewall is in itself insufficient as a means of attempting to access the application layer, as it is then necessary to create an evidence trail of exploits that cannot be hidden from the IPS and anti-malware packet inspection layers. The proposed model is particularly suited to architectures that have a requirement to remain flexible for future scaling (which is often a driver for the adoption of cloud infrastructure), such as those built upon microservices. 
Our solution is mapped to the NIST model in order to assist cloud and IoT system architects to incorporate this work into their own designs.

\end{document}